\documentstyle[11pt,fleqn,leqno,epsfig]{article}
\begin{document}
\title {On the Origin of Cosmic Rays}
 \author{R.Plaga\thanks
        {email:plaga@mppmu.mpg.de}
     \\Max-Planck-Institut f\"ur Physik (Werner-Heisenberg-Institut)
     \\ F\"ohringer Ring 6
     \\D-80805 M\"unchen, Germany}
\maketitle
\markboth{The Origin of Cosmic Rays}
{}

\begin{abstract}
\noindent
The problem in identifying the sites of origin 
of Galactic Cosmic Rays (CRs) is reviewed.
Recent observational evidence from very-high energy 
(VHE, energies above 100 GeV) $\gamma$-ray measurements
is in contradiction with the
surmise that synchrotron radiation from relativistic electrons is
indicative for hadron acceleration. It rather  
points to a CR-acceleration efficiency of 
supernova remnants (SNRs\footnote{
The term ``SNR'' is below always used in the sense: ``the well
known non-relativistic ejecta of supernova with
a typical kinetic energy of 10$^{51}$ ergs''}) 
below one percent, much less than  
the value required if these objects were to be the main sources of 
Galactic cosmic rays (about 30$\%$).
Observations of CR anisotropy and the 
emission of low-energy (energy $<$ 10 GeV)
$\gamma$-rays from the Galactic disk indicate that
the sources of low-energy cosmic rays are distributed
with a Galactocentric radial scale length in the order of 25 $\pm$ 10 kpc, 
much larger than expected if SNRs are the main sources of CRs.
\\
These two facts -
together with the body of evidence from
CR isotope abundances - strongly
suggest that a new class of astrophysical
objects - distinct from SNRs and located manly in
the outer reaches of our Galaxy - is the major source of hadronic CRs
in our Galaxy.
\\
The basic observational features of 
ultra high-energy (UHE, 
energy $>$ 10$^{19}$ eV) CRs are most naturally understood
if the same 
CR sources accelerate CRs up to the highest observed CR energies.
Proposals for the nature of a
new source class are mentioned.
The origin of CRs is still as much shrouded in mystery as it was  
in 1957, when Philip Morrison wrote
a seminal review about CR origin. The potential for discoveries is thus great.

\end{abstract}
\pagebreak
{\it\noindent Thus in the 40 plus years since the publication of Morrison's seminal 
paper\cite{morrison} ... we have come no closer to a definitive model
of cosmic-rays origin. \\
{\tiny Trevor Weekes, August 1999\cite{weekes}}.}
\\
{\it\noindent Until recently the hypothesis that cosmic rays originate in the flashes
of supernovae has been based solely on energy considerations - evidence
that is still far from adequate. \\ 
{\tiny Vitaly Ginzburg, July 1953\cite{ginz53}}.} 
\\
%
\section{Introduction}
\subsection{The problem in historical perspective - outline}
Around 1930 Millikan,Cameron\cite{milli1} and Regener\cite{rege} 
realized
that the energy density of cosmic rays in space is about as high
as that of integrated star-light. This made clear that
cosmic rays are an important cosmic phenomenon, and two questions
were raised.
\begin{figure}[ht]
\vspace{0cm}
\hspace{0cm} \epsfxsize=12.1cm 
\epsfbox{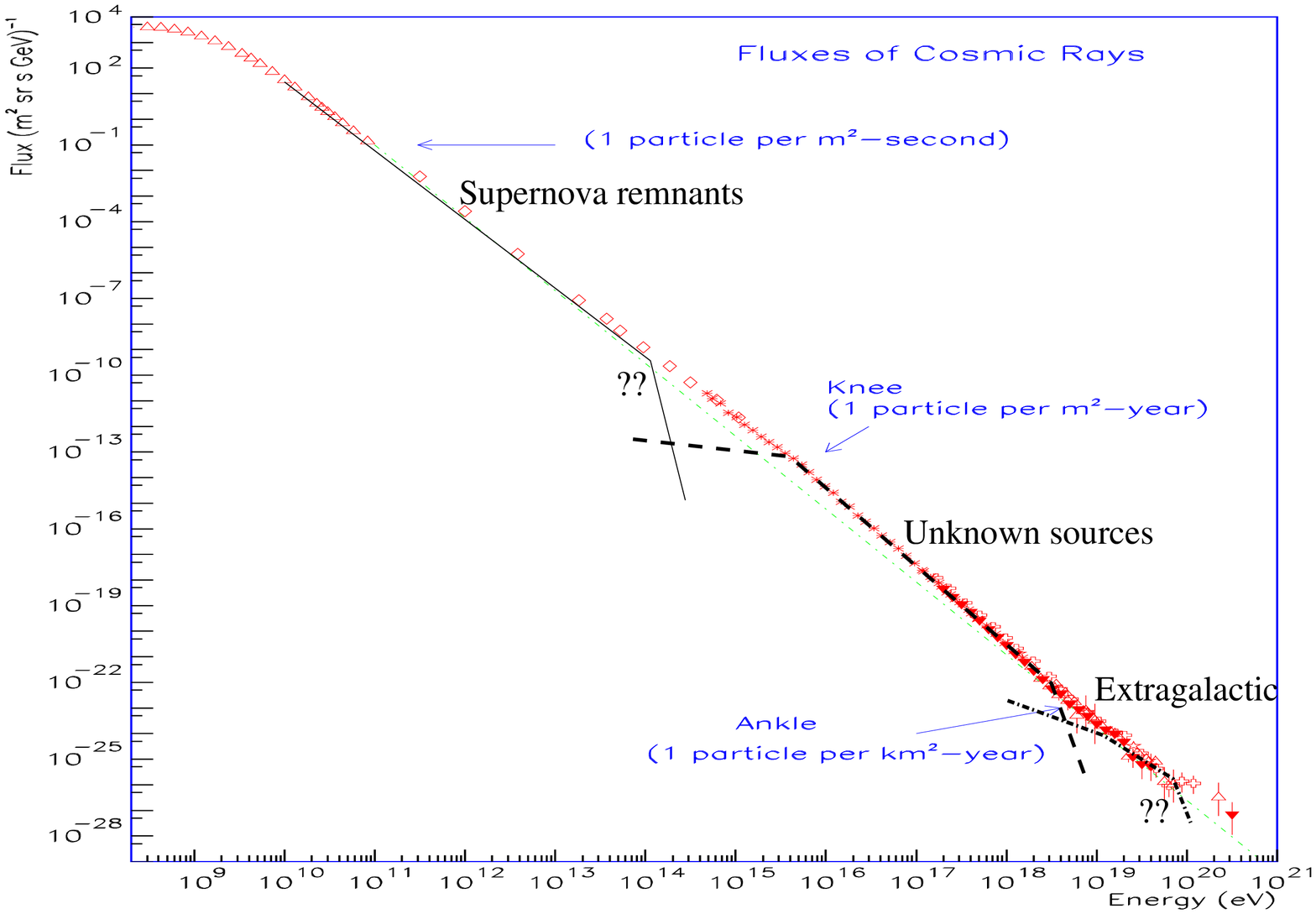}
\vspace{0cm}
\caption{
The flux of hadronic CRs rays as measured by various
experiments as a function of total energy. 
The faint dashed line
through all the data is a power law with a constant index of -3,
and demonstrates that a single power-law is an excellent 
first-order approximation to the experimental data over the whole
energy range.
The deviation from a power law at energies below 10 GeV is due to solar and
terrestrial modulation effects.
The basic components of the experimental CR spectrum
in the ``eclectic (many source classes)'' scenario,
generally accepted presently,
are sketched.
More than four decades have passed since P. Morrison
suggested a slightly different
``hierarchical''
framework\cite{morrison}, in which cosmic-rays fill increasing
shares of the universe with increasing energy.
From the lowest energies up to the ``knee'' at 4 $\times$ 
10$^{15}$ eV SNRs are the main contributors (full line).
Theoretical arguments and experimental evidence suggest an
earlier cutoff around 10$^{14}$ eV, this is symbolized together
with the question marks.
At higher energies up the the ``ankle'' at about 5 $\times$ 
10$^{18}$ eV another source class of unclear
nature, labeled here with ``Unknown sources'' (dashed line) takes over.
At still higher energies an extragalactic component (dash-dotted line) 
dominates, that is expected to cut off at the so-called ``Greisen limit'' of
5 $\times$ 10$^{19}$ eV (the fact that the observed spectrum shows
no evidence for such a cut-off is symbolized by the question marks).
The missing cut-offs and exceptionally
smooth joints of the three components 
(in particular at the ``knee'') finds no
{\it natural} explanation at the moment.
The idea that this theoretical patchwork does not do justice
to the impressing simplicity and unity of the experimental
data strongly suggests itself.
It seems more reasonable to posit that the ``Unknown sources''
component dominates the hadronic CR luminosity at all energies.
(Figure modified from an original of S.P.Swordy)}
\label{fig0}
\end{figure}
\\
1. By which mechanism are cosmic rays accelerated?
\\
2. Where and in which sources does this occur?
\\
In this manuscript I try to summarize what is presently
known about the answer to the {\bf second} question. 
I strive for brevity and will review neither the theoretical
nor the experimental literature exhaustively. 
My style is inspired by a memorable
article of  Philip Morrison with almost
the same title in 1957 \cite{morrison} 
(``On the Origin{\bf s} of Cosmic Rays''
),
because our situation concerning the {\bf second} question 
is hardly better than
the one described
by Morrison like this: ``{\it the broad road (towards the answer)
is finally visible,
but the details may well be all wrong}\cite{morrison}''.
\\
Up to 1948 both questions remained a
complete riddle, revealing a deep fundamental lack of knowledge of
the physical world. Highly
speculative origin theories 
- without a firm basis in contemporary physics - were proposed.
Millikan\cite{millikan} suggested a new fundamental production process and
Lema\^itre\cite{lemaitre} 
explained cosmic rays as relics of the early universe.
These ideas 
stood at the cradle of modern nuclear astrophysics and
cosmology but did not really advance cosmic-ray physics.
Two publications by Baade and
Zwicky from 1934\cite{baade}, 
suggesting an origin of CRs in extragalactical supernovae,
stand out for their impressing far-sight\footnote{Their
basic idea was that relativistic particles with a total energy
between 10$^{53}$ and 10$^{54}$ ergs would be ejected in a supernova.
These particles indirectly ``heat'' the stellar debris
and thus lead to the visible SN event. It might still turn out that this
image is very close to being correct.}.
\\
Then, in a brief, productive
``golden period'' - from about 1948 to 1962 -
Fermi conceived the fundamental theoretical concept for
the acceleration mechanism 
and several crucial 
experimental discoveries provided the basic framework in which we
still grapple today for an answer to the question of the site of origin.
Fermi's brilliantly simple idea was the realization that 
nuclei colliding via magnetic interactions with macroscopic 
gas clouds reach equilibrium only when their energies are
similar, i.e. when the nuclei possess macroscopic kinetic energies.
Henceforth the question was no longer ``how can such high energies
be reached?'' but ``under which circumstances and with what time scale?''
This was the deathblow for the early speculative theories, CR origin had
become a problem of conventional astrophysics.
Since the mid 1980s very speculative 
theories are considered once again, see section \ref{greisen}.
\\
The most important experimental achievement of the ``golden period''
was arguably the determination
of the CR spectrum as a function of total particle energy.
Today
the flux of the dominating 
hadronic component, consisting of protons and heavier nuclei,
of the local non-solar cosmic rays (CRs),
is determined
at energies between about 0.1 GeV and
3 $\cdot$ 10$^{20}$ eV. 
Neglecting solar modulation effects,
the all-particle CR flux   
can be described by power-laws 
with indices in the narrow range between -2.5
and -3.2 over the whole range (more than 10 decades) in energy
(see data points in fig. \ref{fig0}) and a single power law with an index
of -3 is quite a good approximation to the experimental data (see thin
dashed line in fig. \ref{fig0}).
\\
A framework for the explanation
of this spectrum, assuming various
different sites of origin, is at present practically
universally accepted 
and is sketched in fig.\ref{fig0}. 
In sections \ref{low} and \ref{high} I will critically discuss the origin of
the low-energy and very-high energy component, respectively. 
Following a seminal paper by Biermann and Davis section \ref{path}
explains why observations make necessary a departure from the
conventional framework.
Section \ref{proposals} briefly discusses two recent and one historic
proposed alternative major source of CRs. 
In the final section \ref{conclusion} I will give a personal interpretation
of the current situation and lay down some
questions to be answered by future research.
\subsection{Cosmic-Ray Origin: the single most important problem
of ``Astroparticle Physics''}
When I talk to younger colleagues in the field
of astroparticle physics I get sometimes the impression
that for them, the question of {\bf low-energy} cosmic-ray origin is some old
problem left over from the last century.
Nothing could be further from the truth.
Cosmic-ray origin
is the {\bf only} important physics problem
where astroparticle physics is called upon to take the lead
- both theoretically and experimentally -
in providing a full solution for all energies.
In other areas, like the physics of active-galactic nuclei, 
the emission-mechanism of pulsars,
the quest for dark matter or the nature of cosmological background fields
astroparticle physics just supplies some pieces to jigsaw puzzles 
being assembled by others. 
\section{Origin of ``low-energy'' (energies below the ``knee'') CRs}
\label{low}
\subsection{The proposal of SNRs as sources of the low-energy component}
In 1953 Iosif Shklovsky\cite{shklovsky} - armed with his recent identification
of the bright radio source Cas A as the remnant of a historic
supernova explosion (SNR) 
\footnote{An identification that was initially vehemently
resisted by Walter Baade\cite{baade5} who had
proposed SNe as cosmic-ray sources some 20 years earlier.} -
proposed that SNRs (rather than SNe themselves) 
are the main sources of Galactic cosmic rays\footnote{
Shklovsky's priority on this crucial 
idea is clearly stated by V.Ginzburg in an
early review\cite{ginz54}.}.
Shklovsky gave two arguments for his idea:
\\
{ 1.
the total power of the 
of CRs injected into the Galaxy is ``similar'' to the kinetic explosion
power of a Galactic supernovae (i.e. 10$^{51}$ erg/50 years)
\footnote{
\label{proviso}
This ``equality'' does not point 
directly to SNRs as the site of CR acceleration,
rather it only points towards a possible connection ``CRs - supernovae''.
}.
Quantitatively the power of the injected CRs was determined to be
about 30 $\%$  (with an
uncertainty of a factor 3)\cite{fields}
\footnote{This is the most recent value (a previous
careful determination gave the range of 10 - 30 $\%$\cite{dmv}).
In Shklovsky's article the total injected power by SNRs was 
estimated to be about 2 orders
of magnitude smaller
(this was sufficient given the incomplete knowledge of
Galactic CR propagation in 1953).
Such a smaller
efficiency is 
in line with the expectation of Biermann and Davis (see
section \ref{acceff}).
} of the total 
kinetic power of young Galactic SNRs\footnote{Such an argument was first made
made by Baade and Zwicky\cite{baade} in an extragalactic
context. With the same values and assumptions about propagation and
explosion energy as Shklovsky, it was first
made by ter Haar\cite{ter}.}.
\\
2.
SNRs are most prominent non-thermal radio sources because they
produce non-thermal electrons copiously. Shklovsky wrote\cite{shklovsky}: 
{\it ``...it is quite natural
to consider that simultaneously with relativistic electrons, also
occur relativistic heavy particles.}''}
\\
\subsection{A guiding principle: is synchrotron emission indicative of the 
total non-thermal acceleration power?
}
Shklovsky's second point fell on fertile ground. 
In the year before Ludwig Biermann
had made the {\it ``tentative assumption that the radio emission
indicates the order of magnitude of all non thermal emissions.''}\cite{bier52}.
Henceforth, the assumption that the main CR sources are prominent radio sources
was a beacon to most theoretical studies on
CR origin. In 1962 Geoffrey Burbidge formulated the idea
that the other class of celestial radio sources 
that rival SNRs in apparent luminosity, active galaxies
(typified by Cyg A, the second brightest radio source in the sky after
Cas A) are the main sources of all cosmic rays\cite{burb}.
\\
The following decade's discussion
concentrated on a heated controversy
between Burbidge and Vitaly Ginzburg, who had become the
main proponent of Shklovsky's idea\cite{ginz,ginz76}.
A discussion of alternatives (such as e.g. a stellar origin of a very-low
energy component as favored by Morrison\cite{morrison} 
or the ideas of Biermann and Davis\cite{bier58}, discussed in 
section \ref{path}) 
fell completely by the wayside.
\subsection{The present situation: are SNRs the main source of low
energy CRs?}  
In spite of enormous advances in experimental techniques 
in the half century
that has elapsed since the publication of Shklovsky's paper,
{\bf not a single 
additional observational fact directly supporting
the hypothesis that SNRs are the major source of Galactic cosmic
rays has come to light.}
\\
To make matters worse - in what I consider to be the
most important experimental discovery in CR physics since
the 1950s - it was recently learnt that Cas A is {\bf not}
an important source of hadronic cosmic rays (see 
section \ref{acceff} for further discussion). 
This saps the foundations of Shklovsky's second argument.
The beacon mentioned in the previous subsection - that still
guides many\footnote{ 
The gist of a prominent colleague's
only reaction to a recent manuscript\cite{plaga01} was : ``Nobody believes
that because the proposed CR-producing ejecta were not 
seen in the radio.''\label{pbiermann}} -
threatens to have misled us for nearly half a century.
{\bf The dominant sources of hadronic CR are either SNRs or they are not
prominent radio sources.}
The sole direct evidence in favor of a SNR origin of CRs is again
the ter Haar's\cite{ter} energy argument, {\it evidence that is still far
from adequate}''\cite{ginz53}
\\
{\it But if not SNRs, then what}\cite{drury95}?
Let us take a big step back in time and reconsider some 
early arguments {\bf against} SNRs as CR sources
in the light of today's body of experimental evidence,
in an attempt to point a way out of this impasse.

\section{A Path not taken - 1958: Returning to a critique by Biermann and
Davis}
\label{path}
Near the end of the ``golden period'', in 1958, 
two of its principal architects, Ludwig
Biermann and Leverett Davis, summarized some lessons learnt\cite{bier58}.
Their analysis, though certainly noticed\footnote{Their collaboration
on Galactic CRs
was explicitely mentioned in the laudation on occasion of the award of the
Bruce medal to Biermann\cite{lauda}.}, had little impact on the 
consecutive research\footnote{The paper was quoted only five times,
the last time in 1969.}. 
This is a pity since their doubts and some of their surmises 
have been born out by observations in a truly impressive way.
\subsection{
Acceleration efficiency in SNRs} 
\label{acceff}
Biermann and Davis write:
{\it ``With the present frequency of supernovae, it would just be possible
to account for the acceleration of cosmic rays if one assumes a very high
efficiency for the conversion of kinetic to cosmic ray energy.
Even with a higher frequency in the early stages [of Galactic evolution],
the efficiency remains so high (several percent) 
that the present authors find this
proposal difficult to accept.''}
\\ 
Today's value for the required efficiency 
at about 30 $\%$\cite{fields} is even higher. 
Biermann and Davis refrain from telling us why to them
an efficiency of several $\%$ seemed unrealistically high, but
it is easy to guess the reason for this judgment. As
pioneers of solar-system CR physics, they
knew that the efficiency of hadron acceleration 
to relativistic energies in large solar
flare events is typically 0.1 $\%$\cite{bier52}\footnote{
The stress is on {\bf relativistic}, the efficiency for acceleration
to lower, but still non-thermal energies can be much higher\cite{murphy}.}. 
The velocities, matter densities
and magnetic field strengths involved in these events are not radically
different to those in SN ejecta. 
However, there are also important
differences in temporal and spatial scale\footnote{
The acceleration efficiency for relativistic electrons in nova explosions
- intermediate in energy
between solar flares and supernova outbursts - has been
been determined to about 1 $\%$\cite{nova}. This value is
similar to the one in SNRs (section \ref{eorigin}).
Perhaps an efficiency for acceleration
to relativistic energies  
in the order of 0.1 - 1 $\%$ 
is typical for non-relativistic shocks
in a magnetized low-density medium for all species.}, so Biermann and David 
prudently did not
cite solar-system evidence in support of their doubt.
Today's direct evidence on
the {\bf hadron} acceleration efficiency in SNRs
strongly supports their doubts.
\\
Accelerated hadrons interact with ambient matter producing
$\pi_0$ mesons.
The level of very-high energy $\gamma$-ray radiation
expected from $\pi_0$ decay allows to derive the
efficiency of proton acceleration\cite{drury94} and
has been searched for in a considerable
number of SNRs using air-shower arrays and air Cherenkov telescopes.
Let's take a closer look at the two most interesting cases.
\\
\subsubsection{SNR G78.2+2.1 - a cosmic beam dump}
G78.2+2.1 - 
probably the remnant of a core-collapse supernova 
in the Cyg OB9 association 1.4 $\pm$ 0.6 $\times$ 10$^4$ years ago -
is fourth brightest SNR at 1 GHz in the sky\cite{landecker}.
At a distance of only 1.5 $\pm$ 0.3 kpc it subtends an angle of
1 degree, about twice as much as the moon.
Based on various observational indications
Pollock\cite{pollock} suggested that
a bright ``hot spot'' (called ``DR4'') of radio synchrotron
emission in the south-east corner of the SNRs radio shell is due to the dense
molecular cloud Cong 8\cite{cong}, that was recently hit by the advancing
SN shock front.
Since the time Pollock wrote the paper, Fukui and Tatematsu\cite{fukui}
have reconfirmed a CO emission at 2.6 mm 
and Wendker et al.\cite{piepe2} found H$_2$CO absorption
in four out of five positions in DR4.
Because CO and especially H$_2$CO are indicators of high
density, ``{\it ...there is good evidence that interaction with quite
a dense cloud is occurring in this part of the SNR.}
(cited from \cite{piepe2})''. 
The reality of the association between shock wave and molecular cloud
has been doubted
because Frail et al.\cite{frail}
failed to find OH(1720) MHz Maser emission from SNR G78+2.1.
This is unwarranted, Frail et al.\cite{frail} write:''{\it Since masers are
a nonlinear physical phenomenon, depending sensitively on input
parameters the non detection of OH (1720 MHz) maser emission does not
in itself rule out an interaction.''}
In fig.\ref{fig3} I compare the detailed
shape of the IR emitting region at 100 $\mu$m 
as measured by the IRAS satellite (taken from
IRAS ISSA server\cite{iras}) and 
DR4 at 92 cm as measured by the Westerbork
radio telescope (from the WENSS survey\cite{wenss}).
The position and
general elliptical shape, with a major axis from SW to NE,
of the IR and radio emitting region are similar.
This morphological resemblance suggests that
indeed a molecular cloud was heated and compressed 
by the SNR shock wave.
\\
G78.2+2.1 accelerates electrons to a spectrum with a power-law index
of -2.08 $\pm$ 0.04, very close to the theoretically preferred
value of -2.1\cite{zhang}. 
This makes G78.2+2.1 an excellent
candidate for a typical hadronic CR accelerator.
The spectral index
within DR4 is similar to the one in the rest of the remnant\cite{zhang}. 
The entire cloud must then act as a ``beam dump'' 
for accelerated protons. The high matter density in the molecular cloud and
relative closeness of the remnant make DR4 
a most sensitive test-bed of
hadron acceleration in SNRs\cite{ahadr4}.
\\
Fig.\ref{fig2} compares various observational upper limits on VHE emission
from DR4 with theoretical expectation, 
calculated under the assumption that 10$\%$ of the kinetic explosion energy
was converted to hadronic CRs.
The most restrictive limit (from 47 hours
of on-source measurement with the HEGRA system of 
telescopes\cite{hess}) falls below the expected value
by more than a factor of 100.
\subsubsection{Cas A - cradle and bier of the belief that SNRs
are the main source of CRs}
\begin{figure}[ht]
\vspace{0cm}
\hspace{0cm} \epsfxsize=9.1cm 
\epsfbox{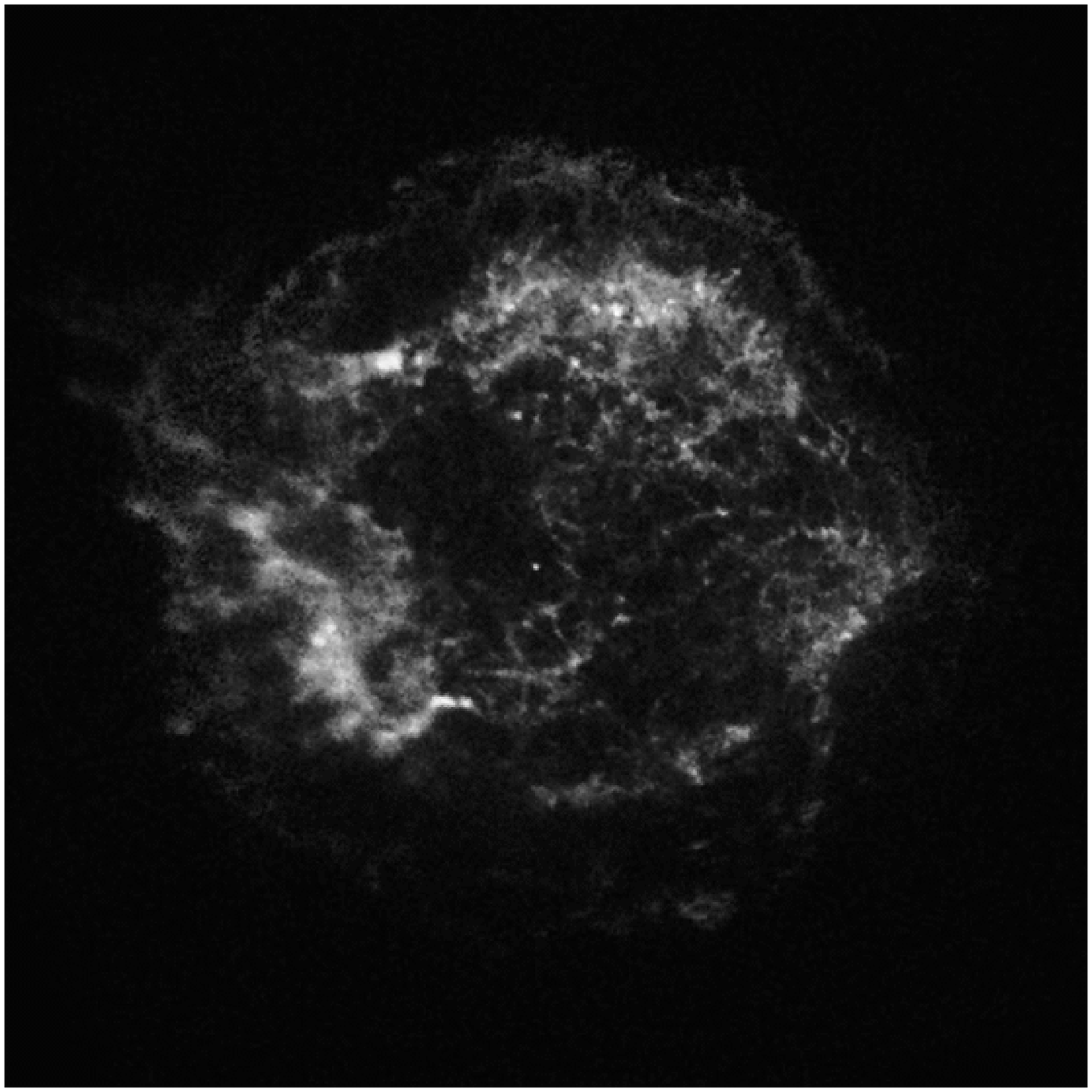}
\vspace{0cm}
\caption{This X-ray image of the Cassiopeia A (Cas A) 
supernova remnant is the official first light image of the Chandra X-ray
Observatory\cite{chandra}. The 5,000 second image was made 
with the Advanced CCD Imaging Spectrometer (ACIS). 
Two shock waves are visible: a
fast outer shock and a slower inner shock. 
The inner shock wave is believed 
to be due to the collision of 
the ejecta from the supernova explosion with a 
circumstellar shell of material, 
heating it to a temperature of ten million degrees Celsius. 
The outer shock wave is progressing into the normal
interstellar medium. 
The bright object near the center may be the 
long sought neutron star or black hole 
that remained after the explosion that produced Cas A.}
\label{fig1}
\end{figure}

\begin{figure}[hbt]
\centering
\begin{tabular}{c@{\qquad}c}
\mbox{
\epsfig{file=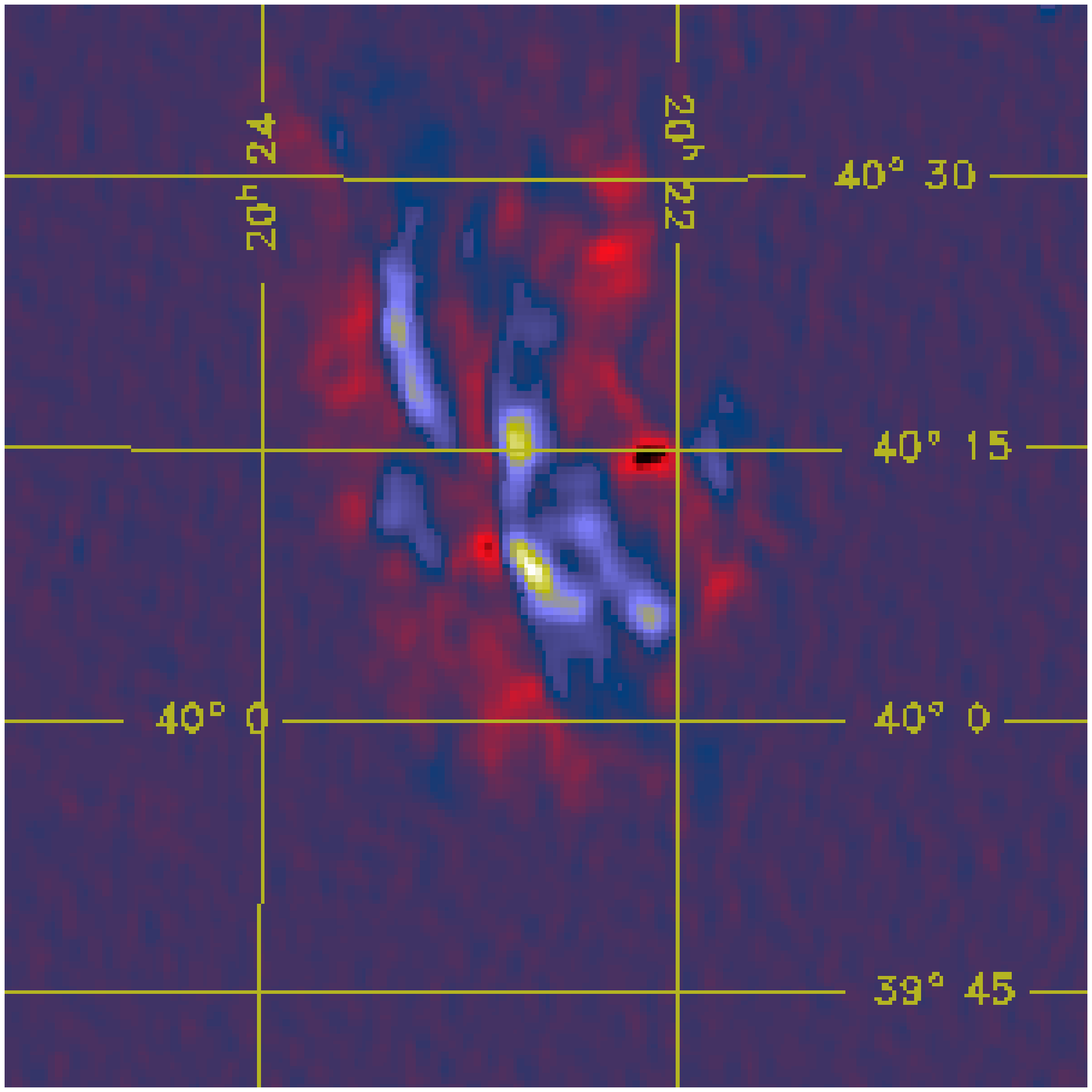 
,width=7cm,height=9cm,clip=,angle=0}}&
\mbox{\epsfig{file= 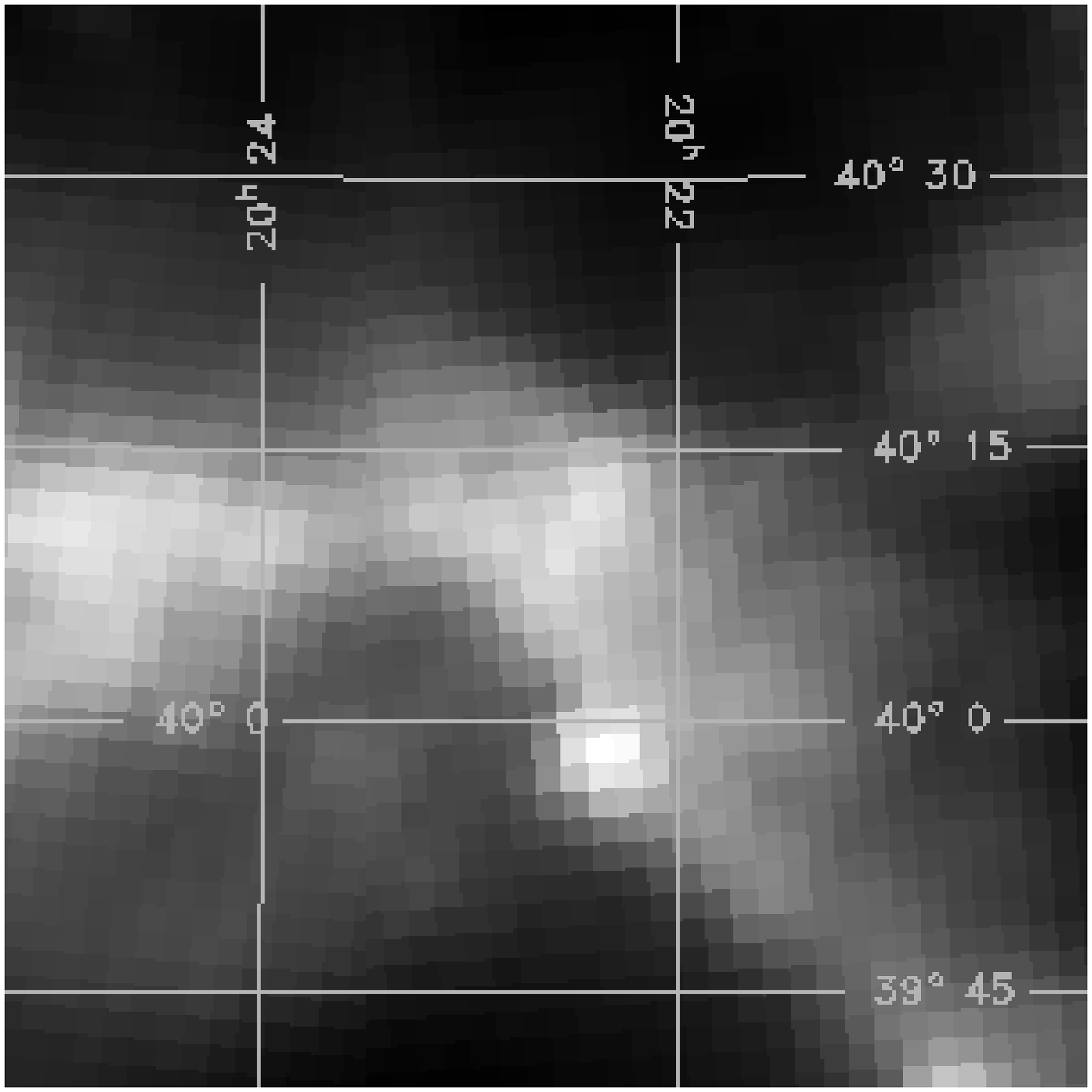
,width=6cm,height=9cm,clip=,angle=0}}
\end{tabular}
\caption{ 
The region around DR4 in the light of 92 cm radio waves
(left image, from the WENSS survey\cite{wenss}) and 100 $\mu$m infrared
light from the IRAS ISSA 
server\cite{iras}. The radio image indicates the distribution of
relativistic electrons emitting synchrotron radiation in the
magnetic field of a molecular cloud compressed
by the SNR shock wave. The infrared emission traces dust
in the cloud heated by the passage of the shock wave.
The similar distribution of both is an argument in Cavour
of an interaction of shock wave and cloud.
Both frames are untreated 1-degree
cutouts from the respective cataloguers.
They are centered on the coordinates
ra: 20h 22m 37.98s; decl: +40d 09m 41.0s (J 2000).
}
\label{fig3}
\end{figure}

\begin{figure}[ht]
\vspace{0cm}
\hspace{0cm} \epsfxsize=11.1cm 
\epsfbox{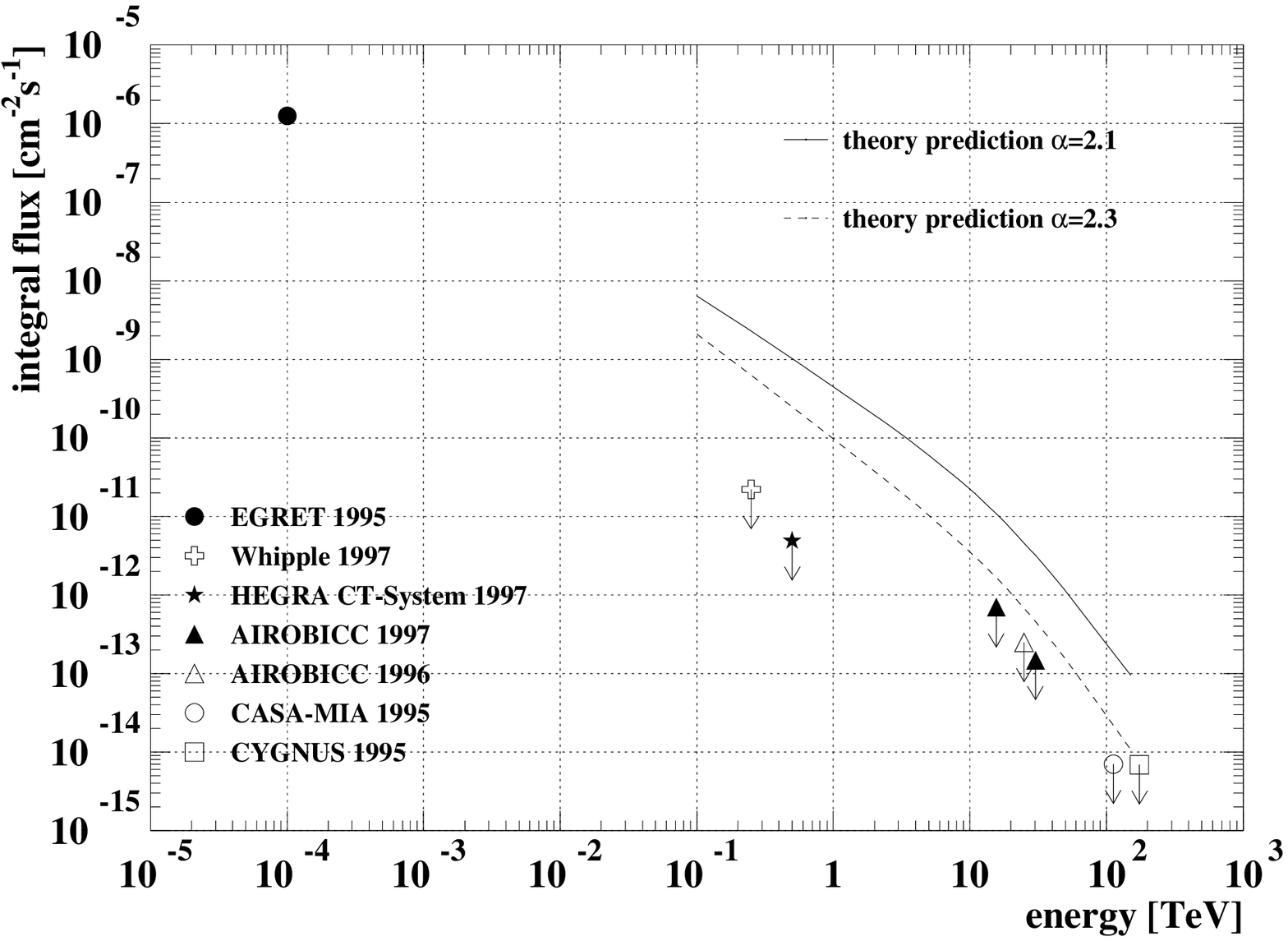}
\vspace{0cm}
\caption{
Various experimental upper limits on the flux of VHE $\gamma$-rays
from the region DR4 in the SNR G78.2+2.1 are compared to a
theoretical 
prediction, derived under the assumption that SNRs are the main
sources of hadronic rays (from the thesis of Prosch\cite{proschthesis}
which contains all the references.)
The predictions under the assumption of two spectral
power-law 
indices of the accelerated hadrons
-2.1 (full line) and -2.3 (dashed line) are shown. 
The universal value for strong shocks
theoretically expected is -2.1. A spectral cutoff at 100 TeV was also
assumed.
The detailed assumptions
used for deriving the theoretical curves are explained in
Prosch et al.\cite{feigl}. This publication contributed the 
point labeled ``AIROBICC 1996''
and was the first to officially announce the failure
of the theoretical predictions by Aharonian et al.\cite{drury94}.
}
\label{fig2}
\end{figure}

Cas A, at a distance of 2.8 - 3.5 kpc, is the remnant of 
Flamsteed's type Ib supernova ``3 Cas'' 
in the year 1680. It is the brightest non-thermal
radio source in the sky and one of the best studied and understood
non-thermal objects\cite{bork}.
It was Cas A that inspired Shklovsky to 
propose that Galactic CRs are mainly accelerated in SNRs 
\cite{shklovsky}.
His idea was refined in the 1970s by models in which the CRs are
mainly accelerated at the outer plane shock front via the first-order
Fermi mechanism\cite{bell,dmv}. Recently very clear experimental evidence from
X-ray measurements for such a shock was found in Cas A (see fig. \ref{fig1}).

Berezhko, P\"uhlhofer and V\"olk\cite{bere} carefully calculated the
expected VHE $\gamma$ flux from Cas A, under the assumption
that CR acceleration efficiency of Cas A is 20 $\%$.
They point out that the observed flux of $\gamma$-rays 
with energies above 500 GeV\cite{puehl}
falls below the expected flux by more than a factor of 100.
The observed signal might well be due to inverse Compton 
emission by CR electrons, that are known to exist in Cas A\cite{puehl}.
The discrepancy would then increase.
\subsubsection{SNRs are not the main sources of cosmic rays - their
efficiency for acceleration is too low}
The upper limits on $\gamma$ radiation from these two remnants 
bound the hadronic acceleration efficiency to
about 20/100=0.2 $\%$. 
This shows that the intuition of Biermann/Davis was correct:
the efficiency with which the kinetic energy of SNRs
is converted to CRs
is not much larger than 1 $\%$, {\bf thus excluding SNRs as the
major source of hadronic Galactic cosmic rays.}
\\
Let us consider various attempts\cite{baring,bere}
to avoid this conclusion.
\\
The different ages of Cas A and G78.2+2.1 exclude 
that the small VHE $\gamma$-ray flux 
is due to a long time scale of hadron acceleration (``late injection'')
or faster than usually expected diffusive CR loss from the remnant.
\\
There is a handful of other SNRs for which less restrictive
or more model-dependent upper limits from 
Cherenkov telescope\cite{heinzi,buckley97,rowell}
and low energy-threshold
air-shower arrays\cite{mizutani,feigl} 
below the theoretical predictions exist.
Therefore speculating 
that only a small fraction of ``special'' 
SNRs are the
main sources of CRs quickly runs afoul with the required efficiency
of particle acceleration that can hardly be much higher than the 30 $\%$
required if all Galactic SNRs contribute equally\footnote{
This is only true of the CR producing SNRs are already known. The idea
of an hitherto unknown class of ``SNRs'' with very different properties 
is indeed plausible and the subject of some proposals presented in section
\ref{proposals}.
}.
\\
The non-detection of $\gamma$-rays in 
the required amount might indeed be 
due to an upper energy cutoff below the threshold
of the detectors (at 200 GeV and higher for various Cherenkov telescopes
and at 3 TeV and higher for various arrays).
Berezhko and V\"olk\cite{bere2} considered
this possibility and aptly
concluded: ``{\it Note however, that in this case SNRs would
hardly be considered as the sources of Galactic CRs.}\footnote{
Because the unity of the spectrum up to the knee implies
a common origin of all CRs below the knee.}''
\\
Finally, the enormity of the discrepancy for the two cases presented
above precludes to invoke a SNR source spectrum steeper
than the one expected from the standard theory 
(power-law index of -2.1). The steepest index
compatible (with difficulty \cite{maurin}) 
with observational evidence from CR abundance ratios is -2.4.
This value still results in a discrepancy from observation
by a factor 10.
\subsection{Anisotropy and Galactic distribution of CRs}
\label{aniso}
\begin{figure}[ht]
\vspace{0cm}
\hspace{0cm} \epsfxsize=9.1cm 
\epsfbox{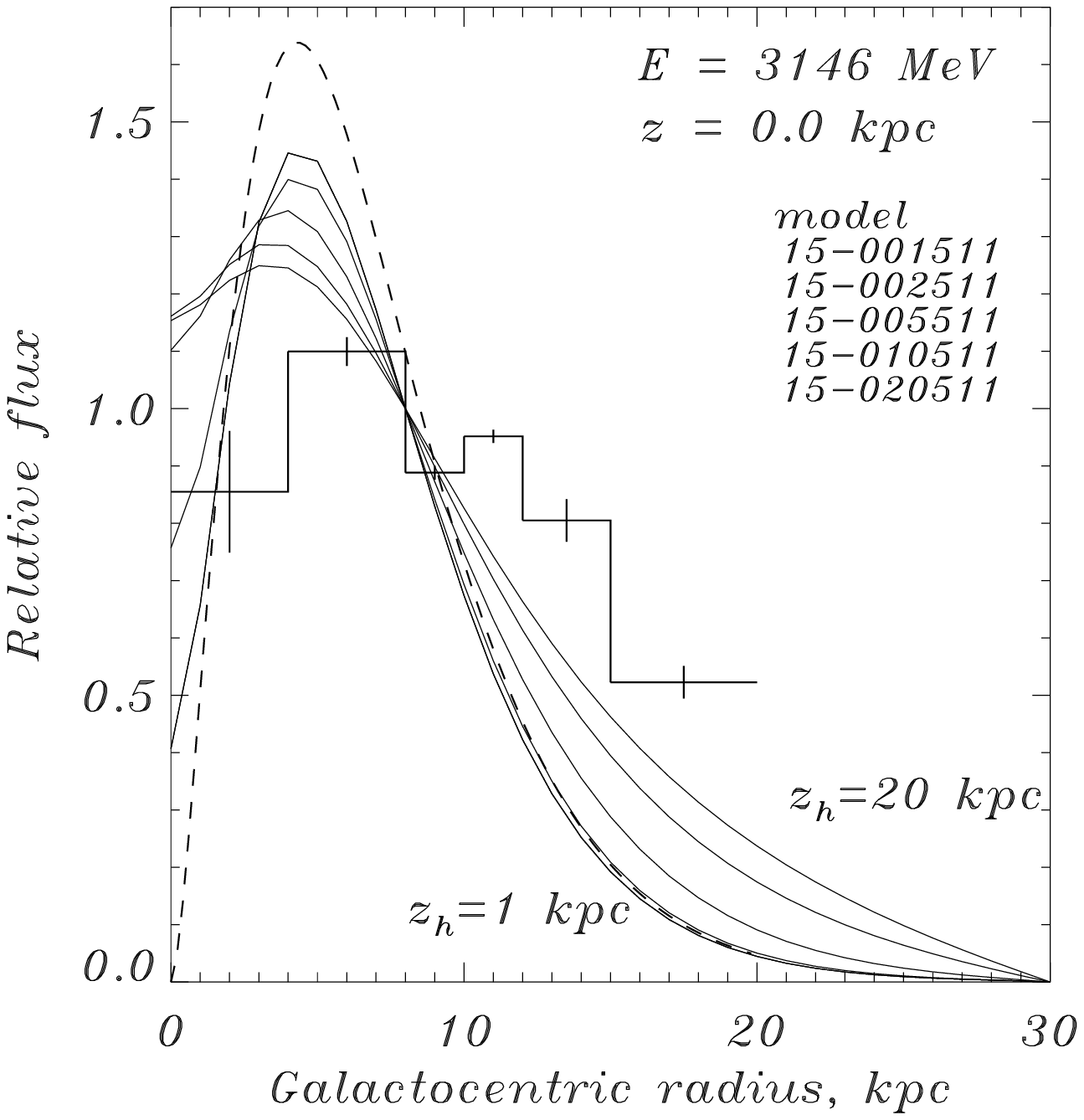}
\vspace{0cm}
\caption{
Radial distribution of 3 GeV protons at $z = 0$,
for diffusive re acceleration model with different 
halo sizes $z_h = 1$, 3, 5,
10, 15, and 20 kpc (five continuous solid curves). The source distribution 
assumed for the calculation of these curves is that for
SNRs as given by Case \& Bhattacharya \cite{case}, shown as a dashed line. The
cosmic-ray distribution deduced from EGRET $>$100 MeV gamma rays
(\cite{mattox96}) is shown as the histogram.
It is seen that the distribution of hadronic cosmic rays is essentially
homogenous up to about 15 kpc. If cosmic rays are mainly produced
in SNRs or similarly distributed objects
a strong concentration towards the Galactic center is expected -
in strong contradiction to the data. (Figure taken from Strong and Moskalenko
\cite{strong1}.)
}
\label{galo}
\end{figure}

Biermann and Davis also directed their attention to
air-shower experiments that had just shown that the amplitude of
anisotropy of Galactic cosmic rays was surprisingly small,
below 10${^{-3}}$ up to 10${^{14}}$ eV and still below 10 $\%$ at 
10${^{17}}$ eV.
These experimental results have stood the 
test of time\cite{linsley83,schmele,clay97,hayashida98}.
\\
If cosmic rays are produced mainly within the solar circle
(as is the case if SNRs are their dominant sources) at earth
their diffusive leakage out of the Galaxy is equivalent to a slow net motion
towards the Galactic anti center.
A finite anisotropy is then expected
due to the so called ``Compton-Getting'' effect\cite{comget}:
less particles are incident in the direction of net motion.
The most recent detailed analysis of this effect 
\cite{ptusk} finds that the amount of anisotropy
expected at an energy of
10$^{14}$ eV is far larger than observed in three out
of four considered models (by factors between 5 and 40).
The remaining model manages to reproduce
an anisotropy of 10$^{-3}$ - but only at the price of implausible
ad hoc assumptions, like a CR source spectrum of the form 
$\sim$ R$^{-2.4}$/(1+(2/R)$^2$)$^{1/2}$ 
(unexpected in any theory) - here R is the rigidity in GV -
and an energy dependence of a cosmic-ray diffusion constant in
flat contradiction to the most recent analyses
of CR abundance ratios \cite{maurin}.
\\
This problem is sometimes dismissed as mere coincidence.
Perhaps the effects of various local sources of CRs conspire to result
in an unusually small anisotropy near earth?
Such a ``conspiracy'' appears unlikely when considering
the determination of the CR density within the Galaxy by using
observations from the 1970s onwards
of low-energy $\gamma$-rays.
The small anisotropy turned out to be the
first indication of a deep lying problem.
\\
The radial scale length of the cosmic-ray density
- derived in the interval between 5 and 20 kpc - has
a value between 16 kpc and 34 kpc in various recent determinations
based in data obtained with the EGRET experiment\cite{mattox96,erlykin96,
huntero97}(earlier determinations based on SAS and COS-B data had
given similar results).
The space-density of SNRs is maximal at 
a distance about 4 kpc from the Galactic center
and falls off with a scale length of about
5 kpc with increasing Galactocentric distance. 
Calculation of CR propagation based on this distribution
predict a Galactocentric gradient which is about 3 - 5 times
smaller than the observed one (see fig.\ref{galo}).
The distribution of CRs within the Galaxy is much more homogeneous
than expected, and this leads to the smaller than expected
isotropy.
There are two possible explanations for the discrepamcy. Either the 
Galactic-propagation properties
or the source location of CRs are different from what was thought
plausible up to now.
\\
With remarkable foresight Biermann and Davis had
anticipated the problem just discussed from the limited
data available to them and proposed an age of the observed 
cosmic rays of several billion years -
one to two orders of magnitude larger than 
normally assumed. 
The longer diffusion time would indeed 
quantitatively result in the observed more
homogenous distribution of CRs in the Galaxy.
However,
their proposal was ruled out by later experimental results:
In particular, the work on the relatively large
concentration of unstable isotopes with lifetimes in the
order of 10 million years
(Be$^{10}$ and Al$^{26}$) - performed after 1958 - has convincingly
proven that the CR ages cannot exceed several hundred million years.
An alternative modification of the propagation properties
suggests that the timescale of CR escape from the
Galaxy might decrease towards the Galactic center\cite{breit}.
This reduces the expected Galactocentric CR-density gradient.
However the inferred total CR power of the Galaxy then exceeds the
conventional value, thus requiring unreasonable
acceleration efficiencies in SNRs (section \ref{acceff}).
\\
It is then most natural to blame the Galactic location of
the CR sources for the small Galactocentric CR-density gradient.
{\bf The main cosmic-ray sources are distributed in the Galaxy
with a Galactocentric scale length in the order of 25 $\pm$ 10 kpc.}
Various systematic uncertainties are
still too large to derive the specific form of the source distribution -
it might e.g. be exponential with a scale length of order 25 kpc
or, alternatively, constant up to 25 kpc
with an additional 
Galactic-center contribution of order of 10$\%$ of the total.
\\
The
distribution of SNRs in the Galaxy
has not only be observed directly\cite{case}, but can also reliably be
inferred - yielding similar results - 
from the one of their precursors, massive stars 
in our Galaxy and their brethren, pulsars\cite{erlykin96}.
This firmly excludes that SNRs may be distributed 
in the Galaxy as required by the assumption that they 
are the main sources of CRs
(as had been suggested by Strong and
Moskalenko\cite{strong1}).
There simply is no known class of potential CR accelerators
located as required. 
Even Galactic dark matter (if it exists) is too
strongly concentrated towards the Galactic center.
{\bf The main sources of hadronic cosmic rays must be a completely novel
\footnote{\rm ``Novel'' is meant in the sense ``not yet directly
observed'', the new source-class might well be related to 
 known phenomena, like supernovae.} class
astrophysical objects.}
\subsection{The Origin of CR electrons in SNRs} 
\label{eorigin}
Biermann and Davis continue after their rejection of the hypothesis
of SNRs as the main CR sources:
\\
{\it ``Supernovae might well
maintain the energy and number balance of the relativistic electrons
responsible for the radio radiation of the halos. This is independent
of the overall energy balance of the nuclei, which constitute $\ge$ 99 $\%$
of the cosmic rays.''}
\\
I think that
Biermann and Davis were correct\footnote{See Dar et al.\cite{dar99}
for a different view.}.
In SNR 1006 about 1$\%$ of the remnant's total kinetic energy
was converted to relativistic electrons\cite{ockie}\footnote{
This value was derived from the expressions in Ref.\cite{ockie}
and the experimental flux at TeV energies\cite{tanimori}.}.
If this value is typical for Galactic SNRs, 
diffusive and radiative losses of Galactic CR electrons can
be replenished by SNRs.
Electrons are observed to be
accelerated in SNRs up to the highest energies observed in the ambient
electron spectrum\cite{tanimori}.
Moreover electrons - unlike the hadrons - 
are {\bf not} found to have a homogeneous
distribution in the Galaxy. Rather they are strongly concentrated towards
the Galactic center. 
The radial scale length of the synchrotron emissivity of the Galactic
thick disk is only 3.9 $\pm$ 0.2 kpc - in good
agreement with theoretical expectation 
under the assumption of an origin 
of the CR electrons in SNRs\cite{beuer}\footnote{This conclusion
only holds under the observationally plausible
assumption that the mean Galactic interstellar B-field falls with
a Galactocentric radial scale length much in excess 5 kpc.}.
\\
The excellent agreement between theoretical expectation and
observational results under the 
assumption of a SNR origin of CR electrons 
makes the glaring failures of theory in the case
of nucleons seem all the more alarming.
\section{Origin of ``high-energy'' (energies above the ``knee'') CRs}
\label{high}
\subsection{Morrison's dark question: absence of a cutoff
in the CR spectrum near the knee}
\label{highenergy}
Philip Morrison\cite{morrison} already realized that SNRs cannot 
accelerate CRs to energies above about 10$^{14}$ eV
(the ``knee'' feature in the primary CR spectrum lies at 
a rather higher energy of 3 $\times$ 10$^{15}$ eV). 
His conclusion has since then been 
confirmed in numerous detailed
theoretical studies\cite{cesarsky83}. 
Impressive
observational evidence on the reality of the cutoff\cite{reynolds}
\footnote{Reynolds and Keohane conclude
that in 14 SNRs they studied, only one (none) has a maximum electron energy
above 100 (1000) TeV and that this limit should also apply to nuclei.} make
it unlikely that recent ingenious suggestions (e.g. \cite{bell01}) 
to  circumvent Morrison's limit are realized in SNRs.
Morrison called the question of the origin
of CRs with higher energies ``dark'', a designation that
remains completely appropriate to the present day\cite{hillas84}.
\subsection{Today's second dark question: absence of the Greisen cutoff}
\label{greisen}
The most recent results from the AGASA air shower experiment indicate that
the CR spectrum continues with a differential index of about -2.5 from
an ``ankle'' at 5 $\times$ 10$^{18}$ eV to  3 $\times$ 10$^{20}$ eV without
any evidence for a spectral cutoff above  5 $\times$ 10$^{19}$ eV that
is expected if UHE CRs are extragalactic\cite{takeda}.
The newest data\cite{agasaspec} favor a power law 
from 5 $\times$ 10$^{18}$ eV
to 3 $\times$ 10$^{20}$ eV without any structure.
Three basic explanations have been brought forward to understand
this\cite{watsonagano}\footnote{
I do not list the possibility of an origin in the Galactic disk
that had been resurrected by some\cite{blasi}.
It requires an iron dominated composition above 10$^{19}$ eV that was
experimentally ruled out recently\cite{ave}.
}:
\\
1. Magnetic fields with a strength 
commonly encountered only in the center of large
clusters unexpectedly exist also in extra-galactic space within about 
10 Mpc of the Milky Way. This allows to ``wash out'' 
both anisotropies and the Greisen cutoff
to a smooth increase of the spectral index 
in the mentioned energy range
\cite{bierb}. UHE CRs could then be due to active galaxies\cite{bira}
\footnote{In addition, a density ratio of CR producers inside
and outside the local supergalaxy higher
than the one suggested by independent astrophysical 
evidence has to be assumed\cite{angela}.
It is obvious that this proposal was born of distress.}.
\\
2. New physics occurs\cite{guenti}. 
The ideas proposed have a similar status as the ones of Millikan
and Lema\^itre mentioned in the introduction.
\\
3. UHE CRs have their origin in the Galactic halo\cite{bur55,hillas,darplaga}.
\\
I concluded in section \ref{aniso} that the sources
of low-energy CRs are a novel type of astrophysical object that is
located mainly in the outer reaches of our Galaxy. The simplest
point of departure for further research is then to adopt option
3 and to postulate that {\bf low and UHE CRs - and then indeed
CR of all energies - have a common origin in the same
novel source class.} 
A ``numerical coincidence'' 
is in favor of such a paradigm. 
Magnetic confinement to our Galaxy is usually expected to
set in at energies below a few times 10$^{18}$ eV.
The ``ankle'' is an increase
in the spectral index of the primary CR spectrum by about 0.5 
at about this energy.
If only one source class dominates CR production at all energies,
this coincidence can easily be naturally understood as the onset of  
Kraichnanian\cite{cesarsky} Galactic confinement\cite{tkac,plaga98}.
\section{Alternatives to SNRs as CR major sources}
\label{proposals}
To my knowledge there have been only two proposals about an
origin of the {\bf bulk} of cosmic rays proposing objects different
from SNRs or pulsars\footnote{The known population of 
Galactic pulsars falls
short by about an order of magnitude in power to replenish the loss
of Galactic CRs, even if high acceleration efficiencies are invoked.}
within the last decade. In addition
there is one historic proposal for an origin of CR in the Galactic halo.
\subsection{Burbidge}
In 1955 Geoffrey Burbidge\cite{bur55} proposed that 
CRs are accelerated up to the
highest energy in the Galactic halo. Density irregularities with
typical velocities of 200 km/sec were proposed as scattering centers.
The ultimate energy source was imagined to be Galactic rotation tapped via
turbulent hydromagnetic frictional effects.
\subsection{Dar and Plaga}
Gamma-Ray Bursts (GRBs) are non-thermal objects that 
are ill understood\cite{fishman}. We are sure that sometimes GRBs
occur in our own Galaxy, but where, how often and with what
kind of remnants remains shrouded in mystery.
It is then an obvious speculation to identify
the main source of CRs with GRBs (section \ref{aniso}).
This has been first proposed by Arnon Dar\cite{dar98b} 
and worked out by Dar and Plaga\cite{darplaga}.
In the latter paper GRBs are proposed as ``universal'' (dominating
at all energies) CR sources.
An origin of CRs in GRBs (at energies above the ``knee'')
was first proposed by Milgrom and Usov\cite{grb4}, 
Vietri\cite{grb2,vietri2} and Waxman\cite{grb3}.)
More recently Plaga\cite{plaga01} proposed that 
bipolar supernovae\cite{bipolar,cen} eject plasma clouds
into the Galactic halo. These ``plasmoids'' 
are posited as the main sources of CRs.
He based this scenario on the ``cannonball model''
of Dar and de R\`ujula\cite{darrujula,darrujula2}.
\subsection{Dermer}
It might be that Dar and Plaga\cite{darplaga} fell into the same trap as
Shklovsky\cite{shklovsky} and Burbidge\cite{burb} before: 
again the source of CRs is directly identified with
a bright non-thermal object in the sky.
Charles Dermer recently suggested\cite{dermer2000} that
- in a addition to GRBs - a special type of stellar
collapse event resulting in ``fireball transients'' (FT) 
occurs in the Galaxy.
FTs are explosion events where $\ge$ 50 $\%$ of the kinetic energy
of the ejecta resides in relativistic ($\Gamma$ $>$ 2) outflows.
These outflows are proposed as
the main source of Galactic CRs. GRBs would then only be the
exceptionally baryon poor ``tip of the iceberg'' of the underlying
novel  FT source class, that is up to now mainly seen in the ``light''
of hadronic CRs. In Dermer's
scenario UHE CRs are produced on extragalactic GRBs, whereas Dar and Plaga
propose Galactic GRBs also for this component.
\subsection{General constraints on alternatives}
I expect that the final answer 
will involve two ingredients:
an origin mainly (but probably not exclusively)
in the Galactic halo (taken from Burbidge and Dar/Plaga) and
a process with a more 
direct and effective conversion of gravitational
into kinetic energy than the one that shaped SNRs 
(taken from Dermer and Dar/Plaga).
This leaves room for ideas quite different from the ones
presented in this section, like an origin in a novel, primordial class of
neutron or quark stars in the Galactic halo.

\section{Conclusion - present status and future progress. A personal view}
\label{conclusion}
\subsection{Present status - my view}
\label{myway}
Summarizing, SNRs are not the main sources of
Galactic cosmic rays. SNRs are
the main sources of one of its minor components, the electrons.
The distribution of the actual main sources of Galactic cosmic rays
differs from that of any other known class of potential CR accelerators.
It stands to reason to speculate that these sources accelerate the
complete CR spectrum up to the highest energies.
The astroparticle physics 
community holds in its hands solid evidence for a completely
novel type of astrophysical phenomenon.
It must rise to the challenge to study this phenomenon,
not by merely waiting for a serendipitous observational discovery
but by systematic theoretical work on alternatives to a SNR origin
of CRs.
\subsection{Present status - prevailing view}
The extremely disappointing experimental situation is
not denied in the astroparticle/cosmic-ray community. It is generally 
and clearly stated
that a CR origin mainly in SNRs is not certain\footnote{
E.g. Matthew Baring skilfully avoids a personal endorsement   
in his summary of the 1999 26th ICRC\cite{bari}
(``... has led to the almost universal acclaim that SNRs are
the site of CR acceleration. Yet ...'')}. 
However, ``{\it faute de mieux}'' (``for lack
of something better'', quote from a report 
by a working group\cite{drury01} 
on ``tests of Galactic cosmic-ray source models'') 
any consideration of alternatives to SNRs as major CR sources is 
refused (not a {\it single} alternative to a SNR origin of
the low-energy component is mentioned
\footnote{
The strongly related concept of super bubbles - agglomerates
of several old SNRs - as a possible site of CR origin\cite{lingen} 
is discussed,
but the authors stop short of considering them as 
the primary CR source class.}). Thus the prevailing view is that
{\bf theoretical} work on the origin of the low-energy component
must remain restricted 
to the {\bf presently known class of SNRs}\footnote{formed of non-relativistic, 
very roughly spherical, SN ejecta}, because it is {\it ``the (only) one model that
has been worked out in some detail}\cite{drury95}''\footnote{In support 
of my evaluation I quote
from the rejection of an anonymous referee on a manuscript suggesting
an alternative to a SNR origin of CRs\cite{plaga01}.
The referee had demanded an impossibly 
large extension of an already relatively
long manuscript and reasoned: 
``{\it Dr. Plaga argues that 15 A \& A pages is quite a lot,
but he is trying to challenge the cosmic ray edifice that stands upon
uncounted thousands of pages, articles, and conferences!}''. Moreover
in none of two recent reviews of CR 
origin\cite{bari,gaisser} any alternative in the sense
of section \ref{proposals} is mentioned, 
though in both cases such alternatives
were presented at the corresponding conferences.
}. 
\subsection{If the prevailing view continues to predominate...}
...this will have two
unhappy consequences:
\\
{\bf 1. The problem of the
origin of CRs will probably be solved outside the 
astroparticle-physics community, when eventually other
astrophysicists will stumble on the CR source class by chance\footnote{
Because VHE astrophysics lacks sensitive all-sky detectors, it seems
likely that the objects will eventually be discovered in the X-ray.}.
\\
2. The vital chance will be lost to be the first to study
the new source class by means of CR evidence.} 
\\
To illustrate the latter point,
Arnon Dar and I proposed an extremely jetted 
geometry of GRBs (later worked out
in detail by him and Alvaro de R\'ujula\cite{darrujula} within
the ``cannonball model'') 
{\bf guided by CR evidence}\cite{darplaga}.
\subsection{Future progress - some important open questions}
Two questions  
are of primary importance for the problem of CR origin 
and serve as a possible point of departure for the research
recommended in section \ref{myway}.  
The third one is of more general interest.
\\
1. There is solid observational evidence that some fraction
of supernovae involves the ejection of ultrarelativistic plasma
with a total kinetic energy larger than the one of SNRs\cite{bipolar,cen}
(the most recent observation of a 
gamma-ray burst\cite{fishman} - supernova association 
is discussed by Lazzati et al.\cite{sngrb}).
\\
{\bf Where are their remnants (observation) and what properties do
they have (theory)
\footnote{A consideration of various anomalous 
non-thermal remnants in our Galaxy
\cite{chaty,shaver,becker,rengarajan} 
might be interesting in this connection.}?
What is the CR total-power and spectrum produced in these events
in our and other galaxies?}
\\
2. There is solid observational evidence\cite{ngc4631,lanzetta} 
that spiral galaxies 
possess the extended, thermal, magnetized halo or 
``corona''
originally  introduced by Spitzer\cite{spitzer}, that our Galaxy forms
no exception\cite{weiner} and that these halos are in a very dynamic,
turbulent state\cite{ngc4631,ngc3079}.
\\
{\bf Where and with what total power 
are particles accelerated in these halos?}
\\
3.
{\bf What is the total cosmic-ray power of our Galaxy?}
\\
If the answer turns out to be significantly higher than
the standard value of 1.5 $\times$ 10$^{41}$ erg/sec\cite{fields}
(as recently advocated by Dar \& De R\`ujula\cite{darpower})
a similar upscale is necessary for other galaxies.
With an increase by more than a factor 10
CR pressure is expected to dominate the thermal 
gas pressure in the center of galaxy
clusters\cite{biermi}.
This would refute one of the strongest arguments\cite{aguirre} against 
``MOND'', Mordehai Milgrom's modified Newtonian dynamics\cite{mondissue}.
\\
{\bf Acknowledgments}
The author is
supported by a Heisenberg fellowship of the DFG.
I thank J\"urgen Gebauer, Silvia Pezzoni and Julian Schlegl
for reading the manuscript and discussions.


\vskip -2.0 cm
\begin{thebibliography}{xxx}

\bibitem{lauda} Abt H.A., 1967, Award of the Bruce Gold Medal to Professor
Ludwig Biermann, PASP 79,197

\bibitem{aguirre} Aguirre A., Quataert E., 2001, Problems
for MOND in cluster and the LY$\alpha$ forest?, astro-ph/0105184,
acc. for publication ApJ

\bibitem{ahadr4} Aharonian F.,Drury L.O'C, V\"olk H.J., 1994,
GeV/TeV gamma-ray emission from dense molecular clouds overtaken
by supernova shells, A \& A 285,645

\bibitem{puehl} Aharonian F., et al., 2001, 
Evidence for TeV gamma ray emission from Cassiopeia A,
astro-ph/0102391, A \& A in print

\bibitem{ave} Ave M., et al., 2001, Constraints on the Ultra High 
Energy Photon flux using inclined showers from the Haverah Park array,
astro-ph/0110613 

\bibitem{baade} Baade, W., Zwicky, F., 1934, Cosmic Rays from Super-Novae,
Proc. Nat. Acad. Sci., 20,259; Baade, W., Zwicky, F., 1934, Remarks on
Super-Novae and Cosmic Rays, Phys.Rev. 46,76

\bibitem{baade5} Baade, W., Minkowski R., 1954, 
Identification of the radio sources in
Cassiopeia, Cygnus A, and Puppis A, ApJ 119,206

\bibitem{baring}
Baring, M.G., Ellison, D.C., Reynolds, S.P., Grenier, I.A., Goret, P.,
1999,
Radio to Gamma-Ray Emission from Shell-Type Supernova Remnants: 
Predictions from Nonlinear Shock Acceleration Models,
ApJ 513,311

\bibitem{bari} Baring M.G., 1999, Cosmic Ray Origin, Acceleration and
Propagation, astro-ph/9912058v2 (Summary, Rapporteur Vol. 26th ICRC, Salt Lake
City)

\bibitem{becker} Becker R.H., Helfand D.J., 1985, A new class of nonthermal
radio sources, Nat 313,115

\bibitem{bell} Bell A.R., 1978,
The acceleration of cosmic rays in shock fronts I, MNRAS 182,147

\bibitem{bell01} Bell A.R., Lucek S.G., 2001, 
Cosmic ray acceleration to very high energy through the 
non-linear amplification by cosmic rays of the 
seed magnetic fields, MNRAS 321,433 

\bibitem{bere} 
Berezhko, E.G., P\"uhlhofer, G., V\"olk, 2001, 
Gamma-ray emission from Cassiopeia A produced by accelerated cosmic rays,
Proc. 27th ICRC, Hamburg,
p.2473

\bibitem{bere2} Berezhko, E.G, V\"olk, H.J., 1999, TeV Gamma Rays Expected
from Supernova Remnants in a Uniform Interstellar Medium, Proc. 26th ICRC,
Salt Lake City, v4, 377

\bibitem{beuer} Beuermann K., Kanbach G., Berkhuijsen E.M., 1985,
Radio structure of the Galaxy, thick disk and thin disk at 408 MHz,
A \& A 153,17
\bibitem{bier52} Biermann L., 1953, Origin and propagation of cosmic rays,
Ann. Rev. Nuc. Sci. 2,335

\bibitem{bier58} Biermann L., Davis L.,
1958, On the origin of Cosmic Rays during the early part of the Evolution
of our Galaxy,
Z.Naturforschg. 13a,909-915

\bibitem{angela} Blanton M., Blasi P., Olinto A.V., 2001, 
The GZK Feature in our Neighborhood of the Universe, 
Astropart.Phys. 15,275

\bibitem{blasi} Blasi M., Epstein R.I., Olinto A.V., 2000,
Ultra-High Energy Cosmic Rays from Young Neutron Star Winds,
astro-ph/9912240, Ap.J. 533,L123

\bibitem{bork} Borkowski K.J, Szymkowiak A.E., Blondin J.M.,
Sarazin C.L., 1996,
A circumstellar shell model for the Cassiopeia A supernova
remnant, ApJ 466,866

\bibitem{breit} Breitschwerdt D., Dogiel V., V\"olk H.J., 1999,
The diffuse galactic gamma-ray gradient, astro-ph/9908011, Proc. 26th
ICRC, Salt Lake City v4, p.259

\bibitem{buckley97} Buckley, J.H., et al., 1997,
Constraints on cosmic-ray origin from TeV gamma-ray observations 
of supernova remnants, A \& A 329,639

\bibitem{bur55} Burbidge G., 1955, 
Halo of radio emission and the Origin of Cosmic Rays, 
Phys.Rev. 101,906-907

\bibitem{burb} Burbidge G., 1962, The origin of Cosmic Rays,
Prog. Theoret. Phys. 27,999

\bibitem{case} Case G., Bhattacharya D., 1996, 
Revisiting the galactic supernova remnant distribution,
A \& AS, 120C, 437

\bibitem{ngc3079} Cecil G., Bland-Hawthorn J., Veilleux S.,
Filippenko A.V., 2001, Jet- and Wind-Driven Ionized Outflows in the 
Superbubble and Star-Forming Disk of NGC 3079, 
astro-ph/0101010, accepted ApJ

\bibitem{cen} Cen R.,1998, Supernovae, pulsars and 
gamma-ray bursts: a unified
picture, astro-ph/9809022, ApJ 507,L131

\bibitem{cesarsky} Cesarsky C.J.,1980,
Cosmic Ray propagation in the Galaxy,
ARA$\&$A, 18, 289-319


\bibitem{chaty} Chaty S., Rodriguez L.F., Mirabel I.F., Geballe T.R.,
Fuchs Y., Claret A., Cesarsky C.J., 2000, A search for possible interactions
between ejections from GRS 1915+105 and the surrounding interstellar
medium, astro-ph/0011297, to appear in A \& A

\bibitem{chandra}
http://chandra.harvard.edu



\bibitem{clay97} Clay R.W., Donough M.-A.,
Smith A.G.K., Dawson B.R., 1998, Anisotropies and the power requirements
for Galactic Cosmic Rays, 
PASA 15, available at
{\tt http://www.atnf.csiro.au/pasa/15$_2$/}

\bibitem{comget} Compton A.H., Getting I.A., 1935,
An apparent Effect of Galactic Rotation on the Intensity of Cosmic Rays,
PR 47,817-821

\bibitem{cong} Cong H.I.L., 1977, A survey of carbon monoxide in Cygnus X,
PhD thesis, Columbia Univ. 
National Aeronautics and Space Administration. 
Goddard Space Flight Center, Greenbelt, MD.



 
\bibitem{dar98b} Dar A., 1998, Cosmic rays and gamma-ray bursts
from microblazars, astro-ph/9809163


\bibitem{dar99} Dar A., De R\`ujula A., Antoniou N., 1999,
A common origin of all the species of high-energy cosmic rays?,
astro-ph/9901004

\bibitem{darplaga} Dar A., Plaga R., 1999, 
Galactic $\gamma$-ray bursters - 
an alternative source of cosmic rays at all energies,
A\&A 349,259

\bibitem{darrujula} Dar A., De R\'ujula A., 2000, 
A cannonball model of gamma-ray bursts: superluminal signatures,
astro-ph/0008474, A \& A in the press

\bibitem{darrujula2} Dar A., De R\'ujula A., 2000a, 
The Cannonball Model of GRBs: Temporal and Spectral Properties of The Gamma
Rays, astro-ph/0012227, submitted to A \& A

\bibitem{darpower}  Dar A., De R\'ujula A., 2001,
What is the Cosmic-Ray luminosity of our Galaxy?, astro-ph/0007306,
CERN-TH/2000-216, ApJ 547,L33

\bibitem{dermer2000} Dermer C., 2000,
Neutrino, Neutron, and Cosmic Ray Production in the 
External Shock Model of Gamma Ray Bursts,
astro-ph/0005440, submitted to
and rejected by ApJ; Dermer C., 2000, Gamma Ray Bursts, 
Cosmic Ray Origin, and the Unidentified EGRET Sources,
astro-ph/0010564, Proceedings of the
Heidelberg 2000 High-Energy Gamma-Ray Workshop, 
ed. F. A. Aharonian and H. V\"olk (AIP: New York)

\bibitem{dmv} Drury, L. O'C, Markiewicz, V\"olk, H.J., 1989,
Simplified models for the evolution of supernova remnants including
particle acceleration, A \& A 225,179

\bibitem{drury94}
Drury L.O'C., Aharonian F.A., V\"olk H., The gamma-ray visibility of
supernova remnants. A test of cosmic-ray origin., A \& A 287, 959

\bibitem{drury95}
Drury L.O'C., 1995, SNRs as gamma-ray sources, Proc. Padova workshop on
TeV Gamma-Ray Astrophysics, ed. M.Cresti, p.76  

\bibitem{drury01} Drury L.O'C., Ellison D.C., Aharonian F.A.,
Berezhko E., Bykov A., Decourchelle A., Diehl R., Meynet G., Parizot E.,
Raymond J., Reynolds S., Spangler S., Tests of galactic cosmic ray source
models (Report of working group 4 at the ISSI workshop on 
Astrophysics of GalacticCosmic Rays), astro-ph/0106046v1



\bibitem{biermi} En\ss lin T., Biermann P., Kronberg P.P., Wu X.-P.,
1997, Cosmic Ray Protons and Magnetic Fields in Clusters 
of Galaxies and their Cosmological Consequences,  astro-ph/9609190,
ApJ 477, 560

\bibitem{erlykin96} Erlykin A. D.,
Wolfendale A.W., Zhang L., Zielinska M., 1996, 
The gradient of cosmic ray protons in the outer Galaxy,
A\&ASS 120,397


\bibitem{fields} Fields B.D., Olive K.A., Cass\'e M., Vangioni-Flam E., 2001,
Standard cosmic ray energetics and light element production, A \& A 370,623

\bibitem{fishman}
Fishman, G. J.  \& Meegan, C. A. A. 1995, 
Gamma-Ray Bursts,
ARAA 33,415

\bibitem{frail} Frail, D.A., et al., 1996, A survey for OH (1720 MHz) maser
emission toward supernova remnants, AJ 111,1651

\bibitem{fukui} 
Fukui, Y., Tatematsu K.,1988, A CO (j = 1-0) 
Survey of Five Supernova Remnants 
at L = 70$^{\circ}$ - 110$^{\circ}$, in: Supernova remnants and the
interstellar medium, IAU coll. 101,eds. R.Roger,T.L.Landecker,
Cambridge University Press, Cambridge, p.261

\bibitem{gaisser} Gaisser T. K., Origin of Cosmic Radiation,
astro-ph/0011524, Proceedings of the International 
Symposium on High energy Gamma-Ray Astronomy 
(Heidelberg, June 26-30, 2000)



\bibitem{ginz53} Ginzburg V.L., 1953, Supernovae and Novae as sources
of Cosmic and Radio Radiation, Dokl.Akad.Nauk.SSSR 92,1133

\bibitem{ginz54} Ginzburg V.L., 1954, Der Ursprung der kosmischen Strahlung
und die Radioastronomie, Fort.Phys. 1,659

\bibitem{ginz} 
Ginzburg V.L., 1957, The origin of cosmic rays,
Prog. Elem. Particle Cosm. Ray Phys. 4,339

\bibitem{ginz76} Ginzburg V.L.,
Ptuskin V.S., 1976, On the origin of cosmic rays: Some problems
in high-energy astrophysics, Rev.Mod.Phys. 48,161-189


\bibitem{hayashida98} 
Hayashida N. et al. (AGASA coll.),1998,
The Anisotropy of Cosmic Ray Arrival directions around
10$^{18}$ eV, astro-ph/9807045

\bibitem{hess}
He\ss \ M., Search for TeV Gamma-Ray Emission from Supernova
Remnants, 25th ICRC, Durban, 1997, v3, p.229

\bibitem{hillas84} Hillas A.M.,
The origin of Ultra-High-Energy Cosmic Rays, 1984,
ARA \& A, 22,425-444

\bibitem{hillas} Hillas M., 1998, Cosmic rays without end, Nat 395,15

\bibitem{huntero97} Hunter S.D et al.,
1997, EGRET Observations of the Diffuse Gamma-Ray 
Emission from the Galactic Plane,
ApJ 481,205

\bibitem{iras}  IRAS Sky Survey Atlas (ISSA), 
http://www.ipac.caltech.edu/ipac/iras/issa.html




\bibitem{cesarsky83} Lagage P.O., Cesarsky C.J., 1983,
The maximum energy of cosmic rays accelerated by supernova shocks,
A \& A 125,249

\bibitem{landecker} 
Landecker, T.L., Roger, R.S., Higgs, L.A., 1980,
Atomic hydrogen in a field in Cygnus X containing the supernova remnant
G78.2+2.1

\bibitem{lanzetta} Lanzetta K.M.,
Bowen D.V, Tytler D.,Webb J.K., 1995, 
The Gaseous Extent of Galaxies,
ApJ 442,538-568

\bibitem{sngrb} Lazzati D., et al., 2001,  
The optical afterglow of GRB 000911: evidence for an associated supernova?
astro-ph/0109287, A + A in press

\bibitem{lemaitre} Lema\^itre G., 1931, The beginning of the world from the
view of quantum theory, Nat 127,1706
Lema\^itre, G., 1949, Cosmological Application of Relativity,
Rev.Mod.Phys. 21,357


\bibitem{lingen} Lingenfelter R.E.,
Higdon J.C, Ramaty R., 2000, Cosmic Ray Acceleration In
Superbubbles and the Composition of Cosmic Rays, 
astro-ph/0004166

\bibitem{linsley83} Linsley J.,
1983, Spectra, anisotropies and composition of cosmic rays above 1000 GeV,
Proc. 18th ICRC Bangalore 12,135



\bibitem{ockie} Mastichiades A., de Jager O.C., 1996,
TeV emission from SN 1006, A \& A 311,5

\bibitem{maurin}
Maurin D.,Donato F.,
Taillet R., Salati P., 2001, 
Cosmic Rays below Z = 30 in a Diffusion Model: 
New Constraints on Propagation Parameters, ApJ 555,585-596


\bibitem{grb4} Milgrom M., Usov V., 1996,
Gamma-ray bursters as sources of cosmic rays,
Astropart. Phys. 4,365 

\bibitem{mondissue} Milgrom M., 2001, Does MOND follow from the CDM paradigm?,
astro-ph/0110362

\bibitem{milli1} Millikan R.A., Cameron G.H., 1928, New precision is cosmic
ray measurements; yielding extension of spectrum and indications of bands,
Phys.Rev. 31,921

\bibitem{millikan} Millikan R.A., Cameron G.H., 1928, 
The origin of cosmic rays,
Phys.Rev. 32,533;
Millikan, R.A., 1949, The present status of the
Evidence for the Atom-Annihilation Hypothesis, Rev.Mod.Phys. 21,1

\bibitem{mizutani} Mizutani K., Utsugi T., 2001, Constraints on the SNR
origin of cosmic rays from gamma-ray observations in the TeV region,
Proc. 27th ICRC, Hamburg, 2459

\bibitem{morrison} Morrison P., 1957, On the origins of
Cosmic Rays,
Rev.Mod.Phys. 29,235

\bibitem{morri2} Morrison P., 1961, The origin of cosmic rays, Handbuch der
Physik, Band 46/1, 1 (Springer, Berlin).

\bibitem{murphy} Murphy R.J. et al., 1997,
Accelerated particle composition and energetics and
ambient abundances from gamma-ray spectroscopy 
of the 1991 June 4 solar flare, ApJ 490,883


\bibitem{watsonagano} Nagano M., Watson A.A.,
2000, Observations and implications of the ultrahigh-energy cosmic rays,
Rev.Mod.Phys. 72,689-732



\bibitem{plaga98} Plaga R., 1998, 
An extragalactic ``flux trapping'' origin of the 
dominant part of hadronic cosmic rays? A \& A 330,833


\bibitem{plagaokkie} Plaga R.,de Jager O.
\& Dar A., 1999,Are Galactic Gamma-Ray Bursters the 
Main Source of Hadronic Non-Solar Cosmic Rays at all Energies?
Proc. 26th ICRC, Salt Lake City, 4,353 

\bibitem{plaga01} Plaga R., 2001, A possible universal origin
of hadronic cosmic rays from ultrarelativisitc ejecta of
bipolar supernova explosions, astro-ph/0106033


\bibitem{pollock} Pollock, A.M.T., 1985,
The probable identification of two COS-B $\gamma$-ray
sources with molecular clouds compressed by supernova remnants,
A \& A 150,339

\bibitem{feigl} Prosch C. et al., 1996, Search for very high-energy
$\gamma$ radiation from the radio bright region DR 4 of the SNR
G78.2+2.1, A \& A 314,275

\bibitem{proschthesis} Prosch C., 1997, Suche nach
Supernova\"uberresten als Quellen der kosmischen Strahlung
im Energiebereich oberhalb von 14 TeV mit dem
Airobicc-Detektor des HEGRA Experiments, PhD thesis, Technical University
Munich


\bibitem{ptusk} Ptuskin, V.L., Jones, F.C., Lukasiak, A., Webber, W.R.,
2001, Diffusion and nuclear fragmentation of cosmic rays: Choice of 
galactic model, Proc. 27th ICRC 2001,Hamburg, 1947

\bibitem{bira} Rachen J.,
Biermann P.L., 1993, 
Extragalactic ultra high energy cosmic rays, I. 
Contribution from hot spots in FR-II galaxies,
A\&A  272,161  

\bibitem{rege} Regener E., 1933, Der Energiestrom der Ultrastrahlung,
Z.Phys. 80,666

\bibitem{rengarajan} Rengarajan T.N., Verma R.P., 1985, Shock-induced
star formation in G357.7-0.1?, Nat 317,415


\bibitem{nova} Reynolds S.P., Chevalier R.A., 1984, A new type
of extended radio emitter: detection of the old nova GK Persei,
ApJ 281,L33

\bibitem{reynolds} Reynolds S.P.,Keohane J.W., 1999, 
Maximum Energies of Shock-accelerated Electrons in 
Young Shell Supernova Remnants, ApJ 525,368

\bibitem{rowell} Rowell G.P. et al., 2000, Observations
of the supernova remnant W28 at TeV energies, A \& A 359,337

\bibitem{agasaspec} Sakaki et al., 2001, Cosmic Ray spectrum
above 3 $\times$ 10$^{18}$ eV observed with AGASA,
Proc. 27th ICRC, Hamburg, 337


\bibitem{schmele} Schmele D., 1998, 
Suche nach r\"aumlichen Anisotropien der kosmischen 
Strahlung im Energiebereich oberhalb von 20 TeV, 
PhD thesis, University
of Hamburg, available on {\tt http://www.desy.de/~schmele/DOKTOR/}

\bibitem{shklovsky} Shklovsky, I.S., 1953, On the origin of cosmic rays,
Dokl.Akad.Nauk.SSSR, 91,475

\bibitem{bierb} Sigl G., Lemoine M., Biermann P., 1999, High energy
cosmic ray propagation in the local supercluster, astro-ph/9806283,
Astropart.Phys. 10,141-156

\bibitem{guenti} Sigl G., 2001,
Ultrahigh Energy Neutrinos and Cosmic Rays as Probes of New Physics,
hep-ph/0109202 

\bibitem{spitzer} Spitzer, L., 1956, On a possible interstellar galactic
corona, ApJ 124,20


\bibitem{mattox96} Strong A.W., Mattox J.R., 1996,
Gradient model analysis of EGRET diffuse Galactic gamma-ray emission,
A\&A 308,L21

\bibitem{shaver} Shaver P.A. et al., 1985, Two remarkable bright supernova
remnants, Nat 313,113

\bibitem{strong1} Strong A.W.,
Moskalenko I.V., 1998, 
Propagation of Cosmic-Ray Nucleons in the Galaxy, astro-ph/9807150v2, 
ApJ 509,212


\bibitem{takeda} Takeda M. et al. (AGASA coll.),1998,
Extension of the Cosmic-Ray
Energy Spectrum Beyond the Predicted Greisen-Zatsepin-Kuzmin
Cutoff, PRL 81,1163; astro-ph/9807192

\bibitem{tanimori} Tanimori, T. et al., 1998, Discovery of TeV Gamma Rays 
from SN 1006: 
Further Evidence for the Supernova Remnant Origin of Cosmic Rays,
ApJ 497,L25

\bibitem{ter} Ter Haar D.,1950, Cosmogonical problems and stellar energy,
Rev.Mod.Phys. 22,119

\bibitem{tkac} Tkaczyk W., 1995, The Cosmic Ray
Spectrum and irregularities in the Galactic magnetic field,
24th ICRC, Rome, 3,293


\bibitem{grb2} Vietri M., 1995, On the acceleration of 
Ultra High Energy Cosmic Rays in Gamma Ray Bursts, 
ApJ  453,883

\bibitem{vietri2} Vietri M., 1996, 
Coronal gamma-ray bursts as the sources of 
ultra-high-energy cosmic rays?, MNRAS 278,L1

\bibitem{heinzi} V\"olk, H.J., 2001, Supernova Remnants: acceleration
of particles and gamma-ray emission, astro-ph/0105356,
(Proc. Moriond Conf., ``Very High Energy Phenomena in the Universe'', 
January 20-27,2001)

\bibitem{bipolar} Wang L., Howell D.A., H\"oflich P.,
Wheeler J.C.,
 2001, Bipolar Supernova Explosions,astro-ph/9912033, ApJ 550,1030

\bibitem{ngc4631} Wang Q.D., Immler S., Walterbos R., Lauroesch J.T.,
Breitschwerdt D., Chandra detection of a hot gaseous corona around the
edge-on galaxy NGC 4631,
astro-ph/0105541, to appear in ApJ Letters

\bibitem{grb3} Waxman E., 1995, 
Cosmological Gamma Ray Burst and the highest energy Cosmic Rays, 
Phys. Rev. Lett. 75,386

\bibitem{weekes} Weekes, T.C., 1999, VHE astronomy before the new millenium,
Proc. Towards a Major Atmospheric Cherenkov detector VI, ed. B.Dingus et al.,
p.3

\bibitem{weiner} Weiner B.J., Williams T.B.,
1996, Detection of H$\alpha$ Emission
from the Magellanic Stream: evidence for an Extended Gaseous
Galactic Halo ,AJ 111,1156-1163

\bibitem{piepe2} Wendker, H.J., Higgs, L.A., Landecker, T.L., 1991,
The Cygnus X region XVIII. A detailed investigation of radio-contimuum
structure on large and small scales, A \& A 241,551

\bibitem{wenss} The Westerbork Northern Sky Survey,
http://www.strw.leidenuniv.nl/~dpf/wenss/

\bibitem{zhang} Zhang X., Zheng Y., Landecker T.L., Higgs L.A., 1997,
Multifrequency radio spectral studies of the supernova remnant G78.2+2.1,
A \& A 324,461  






\end{thebibliography}
\end{document}